\documentclass[12pt]{iopart}
\usepackage[dvips]{graphicx}
\usepackage{bm}
\usepackage{amssymb}
\usepackage{color}
\usepackage{dcolumn}
\newcommand{\figjov}[1]{\includegraphics[width=0.9\textwidth]{#1.eps}}
\newcommand{\figjovmedium}[1]{\includegraphics[width=0.7\textwidth]{#1.eps}}
\newcommand{\Rfinal}{\mathop{R\mathrm{_{final}}}}

\newcommand{\Rinitial}{\mathop{R\mathrm{_{initial}}}}
\newcommand{\Rhalf}{\mathop{R\mathrm{_{half}}}}
\newcommand{\Imax}{\mathop{I\mathrm{_{max}}}}

\newcommand{\Ip}{\mathop{I\mathrm{_{p}}}}

\newcommand{\Ima}[1]{\mbox{$I\mathrm{_{max}}=#1$ mA}}

\newcommand{\CIS}[1]{\mbox{$CIS=#1$\%}}

\newcommand{\shift}[1]{\mbox{$\delta=#1$\%}}

\newcommand{\eg}{\mbox{\emph{e. g.}}~}
\newcommand{\ie}{\mbox{\emph{i. e.,}}~}

\begin{document}

\title{Electrical current-driven pinhole formation and insulator-metal transition in tunnel junctions}

\author{J. Ventura$^{1,2}$, Z. Zhang$^{3,4}$, Y. Liu$^{3,5}$, J. B. Sousa$^{1,2}$ and P. P. Freitas$^{1,3}$}

\address{$^1$IN,  Rua Alves Redol, 9-1, 1000-029 Lisbon, Portugal}
\address{$^2$IFIMUP, Rua do Campo Alegre, 678, 4169-007, Porto, Portugal}
\address{$^3$INESC-MN, Rua Alves Redol, 9-1, 1000-029 Lisbon, Portugal}
\address{$^4$Department of Optical Science and Engineering, Fudan University, Shanghai, China}
\address{$^5$Department of Physics, Tongji University, Shanghai, China}
\ead{joventur@fc.up.pt}


\begin{abstract}
Current Induced Resistance Switching (CIS) was recently observed in
thin tunnel junctions (TJs) with ferromagnetic (FM) electrodes and
attributed to electromigration of metallic atoms in
nanoconstrictions in the insulating barrier. The CIS effect is here
studied in TJs with two thin (20 \AA) non-magnetic (NM) Ta
electrodes inserted above and below the insulating barrier. We
observe resistance (R) switching for positive applied electrical
current (flowing from the bottom to the top lead), characterized by
a continuous resistance decrease and associated with
current-driven displacement of metallic ions from the bottom electrode into the
barrier (thin barrier state). For negative currents, displaced ions
return into their initial positions in the electrode and the
electrical resistance gradually increases (thick barrier state). We
measured the temperature (T) dependence of the electrical resistance
of both thin- and thick-barrier states ($R_b$ and R$_B$
respectively). Experiments showed a weaker R(T) variation when the
tunnel junction is in the $R_b$ state, associated with a smaller
tunnel contribution. By applying large enough electrical currents we
induced large irreversible R-decreases in the studied TJs,
associated with barrier degradation. We then monitored the evolution
of the R(T) dependence for different stages of barrier degradation.
In particular, we observed a smooth transition from tunnel- to
metallic-dominated transport. The initial degradation-stages are
related to irreversible barrier thickness decreases (without the
formation of pinholes). Only for later barrier degradation stages do
we have the appearance of metallic paths between the two electrodes
that, however, do not lead to metallic dominated transport for small
enough pinhole radius.
\end{abstract}
\maketitle

\section{Introduction}
Tunnel junctions (TJ) are magnetic nanostructures consisting of two
ferromagnetic (FM) layers separated by an insulator (I)
\cite{Moodera}. The magnetization of one of the FM layers (pinned
layer) is fixed by an underlying antiferromagnetic (AFM) layer. The
magnetization of the other FM layer (free layer) reverses almost
freely when a small magnetic field is applied. Due to spin dependent
tunneling \cite{Tedrow_Review} one obtains two distinct resistance
(R) states corresponding to pinned and free layer magnetizations
parallel (low R) or antiparallel (high R). Large tunnel
magnetoresistive ratios of over 70\% in Al$_2$O$_3$ \cite{TMR_sixty}
and 150\% in MgO \cite{MgO_sput,MgO_epit} (currently reaching more than 400\% \cite{MgO_410}) based-tunnel junctions can be obtained, making them the most promising candidates for high
performance, low cost, non-volatile magnetoresistive random access
memories (MRAMs) \cite{TJ_MRAMS}. Current research focus on
replacing magnetic field-driven magnetization reversal by a Current
Induced Magnetization Switching (CIMS) mechanism \cite{Slow,
Berger}. Such goal was recently achieved in magnetic tunnel
junctions \cite{CIMS_TJs,CIMS_TJs2} for current densities
\mbox{$j\sim10^7$ A/cm$^2$}. On the other hand, Liu \emph{et al.}
\cite{CIS_first} observed reversible R-changes induced by lower
current densities \mbox{($j\sim10^6$ A/cm$^2$)} in thin FM/I/FM TJs.
These changes were found of non-magnetic origin \cite{CIS2} and
attributed to electrical field-induced electromigration (EM) in nanoconstrictions in the
insulating barrier \cite{CIS_Deac,CIS_JOV_PRB,CIS_JOV_IEEE_TN}. This
new effect, called Current Induced Switching (CIS) can limit the
implementation of the CIMS mechanism in actual MRAMs, and its
understanding is then crucial for device reliability.

The influence of Current Induced Switching on the behavior of the
transport properties of AFM/FM/NM/I/NM/FM tunnel junctions is here
studied in detail (NM$\equiv$non-magnetic). Resistance switching
under relatively low positive electrical currents ($I\approx20$~mA)
produces a low resistance state due to local displacements of metallic
ions from the bottom electrode into the barrier (thin barrier
state). Applying sufficiently negative currents leads to the return
of the displaced metallic ions from the barrier back into the
electrode and so to an increase of the TJ-electrical resistance
(thick barrier state), virtually displaying a reversible behavior.
No time dependent phenomena were observed after each switching event
\cite{CIS_Relaxacao}, indicating that migrated Ta ions remain in
deep energy minima inside the barrier. We compared the temperature
(T) dependence between 20 K and 300 K of the electrical resistance
of the above junction thin- and thick-barrier states obtained after each
switching ($R_b$ and R$_B$, respectively), observing a smaller R(T)
variation in the thin-barrier $R_b$ state.

Applying large currents ($|I|\gtrsim80$~mA) leads to irreversible
resistance decreases due to enhanced barrier degradation. Successive
switching under large $|I|$ produced a gradual evolution from
tunnel- to metallic-dominated transport due to EM-induced barrier
weakening. This effect initially starts with an irreversible mean
barrier thickness decrease, followed by the establishment of metallic
paths between the two electrodes. However, for small enough pinhole
radius, tunnel still dominates transport (dR/dT$<$0) and only
subsequent current-induced growth of pinhole size leads to a
metallic-like temperature dependence (dR/dT$>$0).

\section{Experimental details}
We studied a series of ion beam deposited tunnel
junctions, with thin \emph{non-magnetic} Ta layers inserted just below
and above the insulating AlO$_x$ barrier. The corresponding complete structure was glass/bottom lead/Ta (90 \AA)/NiFe
(50 \AA)/MnIr (90 \AA)/CoFe (40 \AA)/Ta (20 \AA)/AlO$_x$ (3 \AA + 4
\AA)/Ta(20 \AA)/CoFe (30 \AA)/NiFe (40 \AA)/Ta (30 \AA)/TiW(N) (150
\AA)/top lead. Details on sample deposition and patterning processes
were given previously \cite{CIS_JOV_PRB}. Notice that the studied TJ-structure is similar to that of magnetic tunnel
junctions grown for actual applications with the exception of the additional thin Ta layers, thus making comparisons with the FM/I/FM
system easier. In particular, one can separate interface electric effects related to the particular metal layers bounding the oxide layer, from the spin polarization effects originated from the FM layers. Electrical resistance and
current induced switching were measured with a standard four-point
d.c. method. Temperature dependent measurements were performed (on cooling) in a
closed cycle cryostat down to 20 K \cite{JOV(comparative),JOV_APL}. CIS
cycles were obtained using the pulsed current method \cite{CIS2},
providing the \emph{remnant} resistance value of the tunnel junction
after each current pulse. We briefly describe some of the details
used to measure CIS cycles \cite{CIS_JOV_PRB}. Current pulses
($\Ip$) of 1 s duration are applied to the junction, starting with
increasing pulses from \mbox{$\Ip=0$} (where we define the
resistance as $\Rinitial$) up to a maximum $+\Imax$, in small $\Delta I_p$
steps. The current pulses are then decreased through zero (half cycle, $R\equiv\Rhalf$)
down to negative $-\Imax$, and then again to zero ($\Rfinal$), to
close the CIS hysteretic cycle. To discard non-linear I(V)
contributions, the junction remnant resistance is measured between
current pulses, always using a low current of 0.1 mA, providing a $R(\Ip)$
curve for each cycle. Positive current is here defined as flowing
from the bottom to the top NM electrode. With the above definitions, one defines the CIS coefficient:
\begin{equation}\label{Defenição de CIS}
    CIS =\frac{\Rinitial-\Rhalf}{(\Rinitial+\Rhalf)/2},
\end{equation}
and the resistance shift ($\delta$) in each cycle:
\begin{equation}\label{Defenição de delta}
    \delta
    =\frac{\Rfinal-\Rinitial}{(\Rinitial+\Rfinal)/2}.
\end{equation}

\section{Experimental results}
\begin{figure}
\begin{center}
\figjovmedium{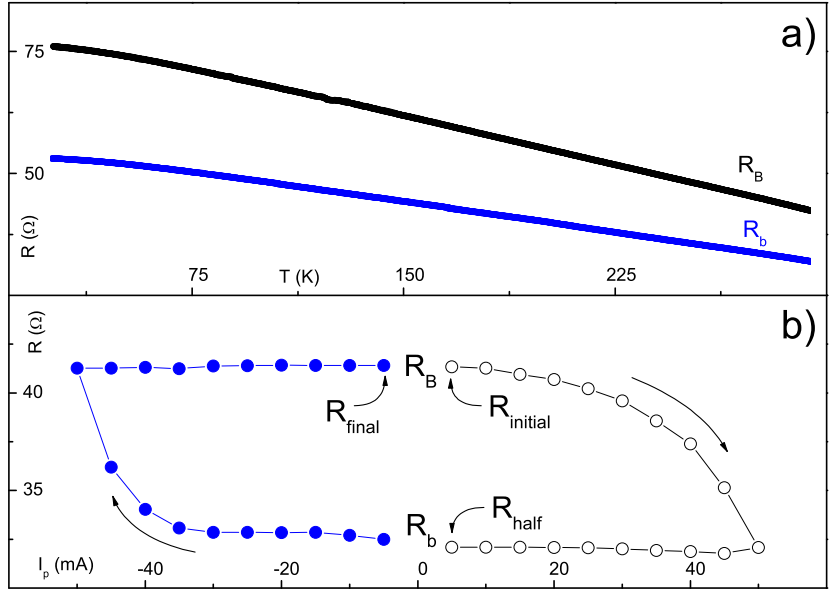}
\end{center}
  \caption{(a) Temperature dependence of the electrical resistance in the low (R$_b$; thin
  barrier) and high (R$_B$; thick barrier) CIS states of TJ1. (b) Half CIS cycle performed with
  \Ima{50} (open circles) that enabled us to change from the (initial) thick
  to the thin barrier state; subsequent half CIS cycle recovering the thick barrier state (full circles)
  and displaying a reversible behavior.}
  \label{fig:R(T)_series2}
\end{figure}

A FM/NM/I/NM/FM tunnel junction with \mbox{$R\approx43~\Omega$} and
R$\times$A$\approx170~ \Omega\mu$m$^2$ (TJ1) was used to study
Current Induced Switching. A CIS cycle spanning pulse currents up to
\Ima{50} is displayed in Fig. \ref{fig:R(T)_series2}(b), giving
\CIS{25} and \shift{-0.1} at room temperature. With
increasing (positive) applied current pulse, switching starts at \mbox{$\Ip\gtrsim20$ mA} through a
progressive (but increasingly pronounced) R-decrease until
\Ima{50}. This switching is associated with local electrical current driven displacement of Ta
ions from the bottom electrode into the insulator
\cite{CIS_JOV_PRB}, decreasing the effective barrier thickness and
so the junction resistance (thin barrier state; R$_b$). Even a small
barrier weakening considerably lowers the tunnel resistance due to
its exponential dependence on barrier thickness. Furthermore, the
switching is asymmetric with respect to the applied current
direction, as observed in previous studies
\cite{CIS_JOV_IEEE_TN,CIS_JOV_PRB} (only ions from the bottom interface are displaced). This effect was related with
the particular sequence of the deposition and oxidation processes during tunnel junction
fabrication \cite{CIS_JOV_IEEE_TN,CIS_JOV_PRB}.

The net atomic flux resulting from an applied electrical field is usually called electromigration. The electromigration force can be divided into two components, one in the direction of the electron flow (wind force) due to the transfer of momentum from electrons to the migrating ions. The other acts in the direction of the electrical field (direct force) and is due to the electrostatic interaction between the electrical field and the direct valence of the ion \cite{EM_Sorbello_book}. Usually, the wind force is much larger than the direct force. However, in our system, the ultra-thin barrier favors intense electrical fields, thus enhancing field-directed diffusion (direct force dominance). Additionally, high applied electrical currents produce large heating and thus thermally enhanced diffusion \cite{CIS_JOV_PRB}. Our results indicate that such diffusion is, in the studied samples, essentially of a local character, ultimately causing pinhole formation (see below).

Returning to Fig. \ref{fig:R(T)_series2}(b), the decrease of $\Ip$ from $+\Imax$ to zero hardly affects the low resistance state. This indicates that displaced Ta ions remain trapped in deep local energy minima inside the barrier. Such low R-state then persists for current pulses down to -35 mA. However, when \mbox{$\Ip<-35$ mA} the (reversed sign) driving force gets strong enough to initiate the return of the displaced ions into their initial positions in the NM layer (causing an increase in R), an effect which is rapidly enhanced until $\Ip=-\Imax$. We then have a thick barrier state (R$_B$), \ie~completely recovering the previous (negative) R-switching which occurred near $+\Imax$. The subsequent change of $\Ip$ from $-\Imax$ to zero, again leaves R essentially unchanged.

We then measured the temperature dependence of the electrical resistance
of the TJ in its \emph{thick} (R$_B$) and \emph{thin} (R$_b$) barrier states
[Fig. \ref{fig:R(T)_series2}(a)]. The resistance of the R$_B$
state steadily increases with decreasing $T$ from \mbox{R$_B=43~\Omega$} at
300 K to \mbox{$76~\Omega$} at \mbox{$T=20$ K} [Fig.
\ref{fig:R(T)_series2}(a)]. Defining the relative R-change between
300 K and \mbox{20 K} as:
\begin{equation}
\alpha=\frac{R_{300 K}-R_{20 K}}{R_{300 K}},\label{Defenição de
alpha}\end{equation} so that $\alpha<0$ ($>0$) indicates tunnel
(metallic) dominated transport, we obtain $\alpha_B=-$78\% for the
thick barrier state. On the other hand, after performing a positive
half CIS cycle ($0\rightarrow\Imax\rightarrow0$) at room temperature
[Fig. \ref{fig:R(T)_series2}(b); open circles], the tunnel junction
was left in its thin barrier state, following the R$_b$(T) measurements. The results now reveal a smaller R-increase
from 300 K down to \mbox{20 K} (R$_b\approx$32
$\Omega$ and 53 $\Omega$, respectively) than in the R$_B(T)$ case, giving $\alpha_b=-66\%$, and so $\alpha_B<\alpha_b<0$.

The R$_B$(T) and R$_b(T)$  curves were fitted to the expressions for two-
and three-step hopping\cite{T-dep-Shang} and phonon-assisted
tunneling\cite{T-dep-CoFeB-Dimop}, revealing a decrease of the hopping
contributions and a slight increase of the phonon-assisted tunneling
when one goes from R$_B$ to the R$_b$ state. This is here attributed to
the decrease of the barrier thickness (reducing the hopping
contribution) and to enhanced excitation of phonons at the
electrodes/oxide interface, by tunneling electrons.

\begin{figure}
\begin{center}
\figjov{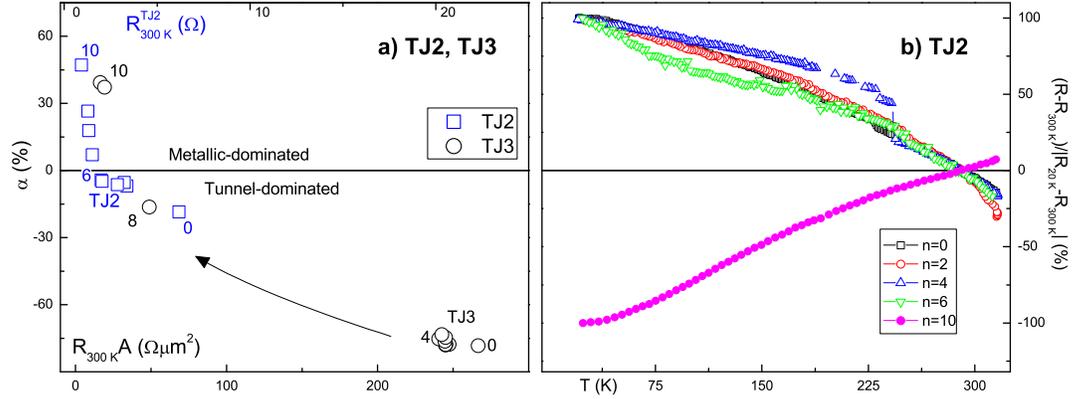}
\end{center}
\caption{(a) The relative resistance change from 300 K to 20 K
($\alpha$) as a function of the $R\times A$ product (lower scale;
for TJ2 and TJ3) and $R_{300 K}$ (upper scale; TJ3 only). Resistance
decrease was induced by EM-driven barrier degradation under high
applied current pulses. The numbers in the Figure indicate how many
CIS cycles were performed before the corresponding $\alpha$-data
point was obtained. (b) Selected normalized R(T) curves in the
300--20 K range (TJ2).}\label{fig:decrease(R)}
\end{figure}

We also studied the effect of increasing barrier degradation
on the (subsequent) temperature dependence of the electrical
resistance of a tunnel junction. For this we performed CIS cycles
with high $\Imax$ (in the 80--110 mA range), each successive one
showing a large negative $\delta$-shift at room temperature
($\Rfinal\ll\Rinitial$), indicating a
progressive and irreversibly barrier weakening \cite{CIS_JOV_PRB}. After each
$n$-th room temperature CIS cycle, we always measured R(T) from
300 to 20 K. No temperature hysteresis was observed, so that on heating back to room temperature, the $n$-th value of the TJ-resistance is essentially recovered. On the other hand, with increasing ($n$) cycling, the junction transport
changed smoothly from tunnel- (dR/dT$<0$) to metallic-dominated
[dR/dT$>0$; Fig. \ref{fig:decrease(R)}(a)]. Two tunnel junctions
were used in this particular study: TJ2 with initial
\mbox{$R=11.3~\Omega$} (\mbox{$R\times A=67.8~\Omega\mu$m$^2$}) and
TJ3 with \mbox{$R=21.6~\Omega$} (\mbox{$R\times
A=259.2~\Omega\mu$m$^2$}).

Both TJs initially ($n=0$; before any CIS cycle) exhibit a
tunnel-dominated R(T) behavior with $\alpha=-20\%$ and $-75\%$, for samples
TJ2 and TJ3 respectively (notice the smaller R$\times$A product
of TJ2). Subsequent CIS cycles with large $\Imax$ led to
irreversible decreases of the TJs resistance and to a steady
increase of $\alpha$ in both samples [Fig.
\ref{fig:decrease(R)}(a)]. Nevertheless, our R(T) data still showed
tunnel-dominated transport (\mbox{$\alpha<0$}) down to
\mbox{R$\times$A$\approx20~\Omega\mu$m$^2$}. The last R(T)
measurement with tunnel-dominated behavior displayed
\mbox{$\alpha=-3\%$} for TJ2 (\mbox{$n=6$};
\mbox{$R\approx2.5~\Omega$}) and \mbox{$\alpha=-15\%$} for TJ3
(\mbox{$n=8$}; \mbox{$R\approx5~\Omega$}). Finally, further
cycling ultimately leads to metallic behavior, with \eg~
$\alpha=47\%$ for TJ2 (\mbox{$n=10$}; $R=0.65~\Omega$) and
$\alpha=40\%$ for TJ3 ($n=10$; $R=1.9~\Omega$), evidencing the
formation of pinholes in the barrier.

We then normalized our R(T) data according to $\frac{R(T)-R_{300
K}}{|R_{20 K}-R_{300 K}|}$. Figure \ref{fig:decrease(R)}(b) displays
selected curves for different stages of barrier degradation (sample
TJ2), obtained after performing the corresponding $n$-th CIS cycle. Although the
shapes of the (tunnel dominated) curves are almost equal, some
display increasing R-steps. On the other hand, the normalized
metallic-dominated transport curves ($n>6$) are all identical.

\section{Discussion} We showed that local atomic displacements cause
irreversible resistance decreases and a progressive change of the
dominant transport mechanism from tunnel to metallic, due to the interplay of two main contributions:
Tunnel through the undamaged part of the barrier (with resistance
$R_t$) and metallic transport through pinholes (resistance $R_m$).
We then write for the measured resistance ($R$):
\begin{equation}
\frac{1}{R}=\frac{1}{R_t}+\frac{1}{R_m}. \label{eq_tunnelemetal}
\end{equation}

We can estimate the evolution of pinhole size with decreasing
TJ-resistance due to barrier degradation. The Sharvin theory
predicts the resistance of a nanoconstriction modeled as a
circular aperture of radius $a$ (between two metallic layers) of
electrical resistivity $\rho$ and electron mean free path $\ell$
\cite{Sharvin_R}. We then have in the ballistic limit ($\ell>>a$):
\begin{equation}
R_m({\textrm{Sharvin}})=\frac{4\rho\ell}{3\pi a^2}.
\label{eq_RSharvin2}
\end{equation}
Since the amorphous Ta layers of the studied TJs have a high
resistivity (\mbox{$\rho\approx150~\mu\Omega$cm})
\cite{CIS_JOV_PRB}, we do not expect the electron mean free
path $\ell$ within a pinhole to be very large (\mbox{$\ell\lesssim a$}). We thus
write the electrical resistance of a constriction in the diffusive
regime ($\ell<<a$), known as the Maxwell resistance
\cite{Wexler_Maxwell_resistance}:
\begin{equation}
R_m({\textrm{Maxwell}})=\frac{\rho}{2a}. \label{eq_RMaxwell}
\end{equation}
A good approximation for the actual pinhole resistance ($R_m$) of a sample with finite $\ell$ is simply
\cite{Wexler_Maxwell_resistance,PRB44_5800,1999PhRvB..60.3963N}:
\begin{equation}
R_{m}=R_m({\textrm{Maxwell}})+R_m~({\textrm{Sharvin}}).
\label{eq_RSharvinplusRMaxwell}
\end{equation}

\begin{figure}
\begin{center}
\figjov{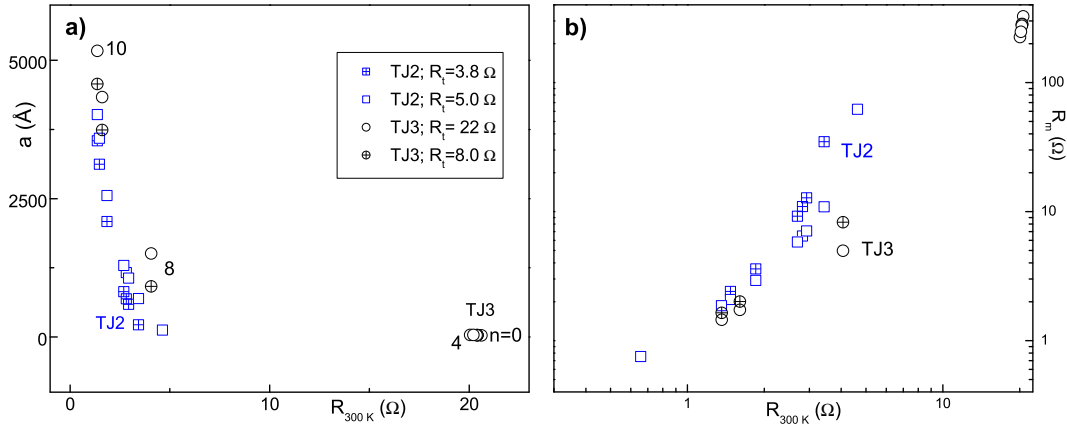}
\end{center}
\caption{(a) Pinhole radius $a$, calculate using Eq.
(\ref{eq_radiusconstriction}), as a function of the experimental
tunnel junction electrical resistance [TJ2 using $R_t=3.8~\Omega$
(open squares) and $R_t=5~\Omega$ (squares with crosses) and TJ3
using $R_t=22~\Omega$ (open circles) and $R_t=8~\Omega$ (circles
with crosses); see discussion for details]. (b) Corresponding
metallic resistance $R_m$, as calculated from Eq.
(\ref{eq_tunnelemetal}).}\label{pinholeArea}
\end{figure}

To calculate the pinhole radius, let us assume that i) a pinhole is
formed just after an irreversible resistance decrease under high
applied current pulses; ii) only one pinhole is formed and grows in
the tunnel junction and, iii) the tunnel resistance remains constant
throughout the successive CIS degrading stages (significant R-variation only arises from the enhancement of the metallic contribution).
Accordingly, $R_t$ is simply the tunnel junction resistance measured
before EM-induced barrier degradation (\mbox{$R_t\approx11~\Omega$} for TJ2 and \mbox{$R_t\approx22~\Omega$} for TJ3). This allows us to estimate the
metallic resistance $R_m$ using Eq. (\ref{eq_tunnelemetal}) and the
measured R-value [see Fig. \ref{pinholeArea}(b)]. Using Eqs.
(\ref{eq_RSharvin2})-(\ref{eq_RSharvinplusRMaxwell}), we can then express the pinhole radius as a function of
$\rho$, $\ell$ and $R_m$:
\begin{equation}
a=\frac{3\pi\rho+\sqrt{3\pi\rho}\sqrt{64\ell R_m+3\pi\rho}}{12\pi
R_m}.\label{eq_radiusconstriction}
\end{equation}

Using \mbox{$\rho\approx150~\mu\Omega$cm} for the Ta layers (and pinholes) and assuming
\mbox{$\ell\approx5$ \AA}, we can then estimate the pinhole radius
[Fig. \ref{pinholeArea}(a)]. Notice that the actual pinhole composition is not entirely known, and the possibility of Ta-Al bonding cannot be excluded. However, the migration of the Ta ions into the amorphous Al$_2$O$_3$ barrier should give rise to a disordered pinhole structure with a high resistivity. For TJ3, as expected, one obtains a  pinhole radius which increases with
decreasing resistance: from about \mbox{30 \AA} for \mbox{$n=4$}
(\mbox{$R_m=250~\Omega$}) to \mbox{1500 \AA} for \mbox{$n=8$}
(\mbox{$R_m=5~\Omega$}), and finally to \mbox{4300 \AA} for
\mbox{$n=10$} (\mbox{$R_m=1.7~\Omega$}). Notice however that this
simple model does not fully describe our data.
In particular, it underestimates the metallic resistance: For
$n=8$ (for which we still observe tunnel dominated
transport; $\alpha<0$) we obtain \mbox{$R_m\approx5~\Omega$}, which
is already smaller than the (assumed constant) tunnel resistance,
\mbox{$R_t\approx22~\Omega$}. Our model then predicts a metallic
R(T) behavior for $n=8$, which contradicts our
data. Notice that such $R_m$ value depends only on
$R_t$ used in Eq. (\ref{eq_tunnelemetal}) and not on $\rho$ or
$\ell$, which are used only to estimate $a$.

We conclude that the initial EM-driven irreversible resistance
decrease is not due to the formation of pinholes [assumption (i)]
but to the progressive weakening of the tunnel barrier (decreasing
barrier thickness) and that the minimum experimental $R$-value
without pinholes ($R=R_t$) is considerably lower than the initial TJ
resistance of \mbox{$22~\Omega$}. Using Eq. (\ref{eq_tunnelemetal}),
we estimate that, to ensure $R_t>R_m$ up to $n=8$ we must have
\mbox{$R_t\approx8~\Omega$}. Then, the initial
$22~\Omega\rightarrow8~\Omega$ R-decrease corresponds only to a
barrier thickness reduction ($\delta t$) without the formation of
pinholes. We estimate \mbox{$\delta t\approx1.3$ \AA}
\cite{CIS_JOV_PRB}.

Assuming then \mbox{$R_t\approx8$ $\Omega$}, we can calculate the
new pinhole radius using Eq. (\ref{eq_radiusconstriction}) [Fig.
\ref{pinholeArea}(a)]. For \mbox{$n=10$} we obtain \mbox{$a=900$
\AA} (\mbox{$R_m=8.3~\Omega$}) and for \mbox{$n=12$}, \mbox{$a=3700$
\AA} (\mbox{$R_m=2.0~\Omega$}), which, nevertheless, are close to
the values obtained above (considering \mbox{$R_t\approx22~\Omega$}).
Furthermore, \mbox{$R_t\approx8~\Omega$} is not the only value that
adequately adapts to the observed $\alpha(R)$ dependence and we
estimate \mbox{$3.5~\Omega\lesssim R_t\lesssim8~\Omega$}. For
\mbox{$R_t\approx3.5~\Omega$} one has pinhole formation only for
$n=11$, already having $R_m<R_t$ (\mbox{$a=2500$ \AA}). In this case
(\mbox{$R_t\approx3.5~\Omega$}), the formation of a pinhole
immediately leads to metallic-dominated transport.

For TJ2, we have more $\alpha$-values near the tunnel/metallic
progressive transition, allowing us to estimate $R_t$ within a
narrower interval. First, notice that (as in the case of TJ3) if we
use the initial TJ2-resistance (\mbox{$R_t=R\approx11~\Omega$}) to
calculate $R_m$, we again obtain $R_m<R_t$ for a R(T) data still dominated by tunneling ($\alpha<0$). Thus,
we again conclude on the initial progressive weakening of the tunnel
barrier (decreasing barrier thickness), leading to the decrease of
the TJ electrical resistance without formation of
pinholes. We predict [using Eq. (\ref{eq_tunnelemetal})] that
\mbox{$3.8~\Omega\lesssim R_t\lesssim5~\Omega$} (Fig.
\ref{pinholeArea}), which corresponds to a barrier thickness
decrease satisfying \mbox{$1.1~$\AA$\lesssim \mid\delta t\mid\lesssim1.3$~\AA}. Using
the two mentioned limiting $R_t$-values, we again observe that $R_m$
decreases with decreasing TJ electrical resistance [Fig.
\ref{pinholeArea}(b)], denoting the increase of pinhole radius [Fig.
\ref{pinholeArea}(a)]. In the case of TJ2, we observe that pinholes
are already formed (\mbox{$a\approx1000$ \AA}) while $\alpha<0$ (demonstrating tunnel dominated
transport; $R_m>R_t$). Further current-induced decrease of the TJ
resistance is seen to be due to the growth of the pinhole size, which
enhances the metallic conductance and
ultimately leads to metallic-dominated transport ($\alpha>0$).

\section{Conclusions}
We showed that the initial insulating barrier
degradation under high electrical currents arises from irreversible
barrier thickness decrease (\mbox{$\delta t\approx-1.3$ \AA}) due to
localized displacement of ions from the electrodes into the barrier,
without the formation of pinholes. Such barrier weakening leads to
higher $\alpha$ values in our R$_b$(T) and R$_B$(T) measurements
(with $\alpha_b>\alpha_B$), suggesting that irreversible and reversible
switching arise from the same physical mechanism. Under adequate
experimental conditions we might even reversibly switch between
$\alpha_B<0$ (tunnel-dominated transport) and $\alpha_b>0$
(metallic-dominated transport), and vice-versa, by local electrical current driven ion diffusion.
Such phenomenon was recently observed by Deac \emph{et al.}
\cite{CIS_Deac} in ultra-thin TJs (barrier thickness
\mbox{$t=5$~\AA}).

Increased barrier degradation leads to the formation of metallic
paths between the two electrodes that, however, \emph{do not lead to
a metallic dominated TJ transport} for small enough pinhole radius. The
increase of such radius gradually leads to the decrease of the metallic
(Sharvin+Maxwell) resistance and thus to the ultimate dominance of
metallic over tunnel transport.

\section*{Acknowledgments}
Work supported in part by FEDER-POCTI/0155,
POCTI/CTM/45252/02 and POCTI/ CTM/59318/2004 from FCT and
IST-2001-37334 NEXT MRAM projects. J. Ventura is thankful for a FCT
post-doctoral grant (SFRH/BPD/21634/2005).

\section*{References}
\bibliography{Biblio}

\end{document}